\documentclass{appolb}
\usepackage{graphicx}
\usepackage{color}
\usepackage{amsmath,amssymb}
\usepackage{wrapfig}

\begin{document}
\title{Universal scaling properties of QCD \\ close to the chiral limit%
\thanks{Presented at workshop on Criticality in QCD and the Hadron Resonance Gas; 29-31 July 2020, Wroclaw, Poland.}%
}
\author{Olaf Kaczmarek, Frithjof Karsch, \\ Anirban Lahiri\thanks{Speaker}, Christian Schmidt
\address{Fakult\"at f\"ur Physik, Universit\"at Bielefeld, D-33615 Bielefeld, Germany}
}
\maketitle
\begin{abstract}
We present a lattice QCD based determination of the chiral phase transition
temperature in QCD with two massless (up and down) and one strange quark
having its physical mass.  We propose and calculate two novel estimators for the chiral transition
temperature for several values of the light quark masses, corresponding to Goldstone
pion masses in the range of $58~{\rm MeV}\lesssim m_\pi\lesssim 163~{\rm MeV}$. The chiral
phase transition temperature is determined by extrapolating to vanishing pion
mass using universal scaling relations. After thermodynamic, continuum and chiral
extrapolations we find the chiral phase transition temperature
$T_c^0=132^{+3}_{-6}$~MeV. We also show some preliminary calculations
that use the conventional estimator for the pseudo-critical temperature
and compare with the new estimators for  $T_c^0$.
Furthermore, we show results for the ratio 
of the chiral order parameter and its susceptibility and
argue that this ratio can be used to differentiate between $O(N)$ and $Z_2$
universality classes in a non-parametric manner.
\end{abstract}
  
\section{Introduction}
By now it is well established that for physical values of light and strange quark
masses QCD undergoes a smooth crossover from a low temperature hadronic phase
to a high temperature partonic phase \cite{Aoki:2006we,Ding:2015ona}. The
chiral crossover temperature has been determined in various numerical
studies of lattice QCD \cite{Aoki:2009sc,Bazavov:2011nk,Bonati:2015bha,Bazavov:2018mes,Borsanyi:2020fev}.
On the other hand the order of the QCD transition in the chiral limit
with two massless degenerate quarks has been a celebrated topic without
any concrete conclusion, yet. It has been argued long back \cite{Pisarski:1983ms}
that \textquotedblleft effective restoration\textquotedblright\ of $U(1)_A$,
which is broken in vacuum, could play a very important role determining the
order of the chiral phase transition for two massless flavors. When the $U(1)_A$
remains broken at the chiral transition temperature, the chiral transition is expected
to belong to the $O(4)$ universality class \cite{Pisarski:1983ms} which is the
most celebrated scenario till date. In case the $U(1)_A$ gets effectively restored
at the chiral transition then the latter may become first order \cite{Pisarski:1983ms},
although second order phase transition belonging to other 3-d universality classes
could also become relevant \cite{Butti:2003nu,Grahl:2013pba,Pelissetto:2013hqa,Sato:2014axa}.
If the chiral transition is first order then there exists an endpoint, belonging to
$Z_2$ universality class, at a non-zero value of light quark mass, $m_l^c>0$ where
the chiral susceptibility will diverge.

In this contribution we present the first lattice QCD based determination \cite{Ding:2019prx}
of the chiral phase transition temperature.
Since there is no direct evidence for a first order phase transition down to a quite
small pion mass, we employ the $O(N)$ scaling to calculate the critical temperature
for vanishing light quark masses. We introduce and present results for two novel
estimators of $T_c^0$ which are reliable even for finite quark masses. We also
present result for the ratio of the chiral order parameter and its susceptibility and argue
that this ratio can be very effective in differentiating between $O(N)$ and $Z_2$
universality classes in a non-parametric manner.

\section{Formalism}
We start with the definition of the quark condensate,
\begin{equation}
\langle \bar\psi \psi\rangle_f = \frac{T}{V} \frac{\partial 
\ln Z(T,V,m_u,m_d,m_s)}{\partial m_f} \; .
\label{pbp}
\end{equation}
In the chiral limit, the light quark chiral condensate
$\langle \bar\psi \psi\rangle_l= (\langle \bar\psi \psi\rangle_u+\langle \bar\psi \psi\rangle_d)/2$,
serves as an exact order parameter for the spontaneous breaking of chiral symmetry at low temperature.
Additive and multiplicative renormalization have been taken care of by introducing \cite{Bazavov:2011nk}
a combination of light and strange quark condensates,
\begin{equation}
M = 2 \left( m_s \langle \bar\psi \psi\rangle_l - m_l \langle \bar\psi \psi\rangle_s
\right)/f_K^4 \; ,
\label{M}
\end{equation}
where $f_K=156.1(9)/\sqrt{2}$~MeV, is the kaon decay constant, used as a normalization constant.
The chiral susceptibility is defined as, 
\begin{eqnarray}
	\hspace*{-0.2cm}\chi_M &=& \left. 
	m_s (\partial_{m_u}+\partial_{m_d}) M \right|_{m_u=m_d} \; .
\label{chim}
\end{eqnarray}

Close to a $2^{nd}$ order phase transition, $M$ and $\chi_M$ are expected to be
described by universal finite-size scaling functions $f_G(z,z_L)$ and $f_\chi(z,z_L)$
\cite{Engels:2014bra},
\begin{eqnarray}
M &=& h^{1/\delta} f_G(z,z_L) + f_{sub}(T,H,L) \; , \nonumber \\
\chi_M &=& h_0^{-1} h^{1/\delta-1} f_\chi(z,z_L) +\tilde{f}_{sub}(T,H,L) \; ,
\label{scalingexpect}
\end{eqnarray}
where the scaling variables in the arguments are defined as
$z=t/h^{1/\beta\delta}$ and $z_L= l_0/(Lh^{\nu/\beta\delta})$,
with $t=(T/T_c^0-1)/t_0$ denoting the reduced temperature; $h=H/h_0$
with $H=m_l/m_s$ being the symmetry breaking field, and $L$ denoting the linear extent of
the system, $L\equiv V^{1/3}$.
The normalization constants $t_0,\ h_0$ and $l_0$ appearing in
definition of the scaling variables are non-universal parameters.
$f_{sub}(T,H,L)$ and $\tilde{f}_{sub}(T,H,L)$ denote 
sub-leading contributions which arise due to contributions from corrections-to-scaling
\cite{Hasenbusch:2000ph,Engels:2000xw} and regular terms, away from the
critical point, for $M$ and $\chi_M$ respectively.

For large enough system sizes, the peak in the scaling function $f_\chi(z,z_L)$ has been
used as the usual estimator for the pseudo-critical temperature $T_p$, which 
scales as
\begin{equation}
T_p(H,L)=
T_c^{0} \left( 1+ \frac{z_p(z_L)}{z_0} H^{1/\beta\delta} \right)\ +\ \text{sub-leading} \; ,
\label{Tpc}
\end{equation}
with $z_0=h_0^{1/\beta\delta}/t_0$.
The universal quark mass dependence of $T_p$ is described by the first term and
\textquoteleft sub-leading\textquoteright\ represents contributions from corrections-to-scaling and regular terms.
In principle, the situation is same for scaling of a temperature $T_X(H,L)$ 
defined at any fixed value $z_X$ ($X=p$ in Eq.~\ref{Tpc}).
Depending on the value of $z_X/z_0$, $T_X(H,L)$ may change significantly within a given window of $H$,
towards chiral limit \cite{Bazavov:2011nk}. This makes the chiral extrapolation complicated due to
increasing importance of the contribution from the sub-leading terms.
Here we consider two estimators \cite{Ding:2019prx,Ding:2018auz} for $T_c^0$, defined close to or at $z=0$,
in the thermodynamic limit resulting in an order of magnitude smaller mass 
variation in Eq.~\ref{Tpc}. 
Pseudo-critical temperatures $T_\delta$ and $T_{60}$ are defined through,
\begin{eqnarray}
\frac{H \chi_M (T_\delta,H,L)}{M(T_\delta,H,L)} &=&\frac{1}{\delta}  \; ,
\label{ratio}\\
\chi_M(T_{60},H) &=& 0.6 \chi_M^{max} \; .
\label{T60}
\end{eqnarray}
Since $z_\delta\equiv z_\delta(0)=0$ and $z_{60}\equiv z_{60}(0) \simeq 0$,
These pseudo-critical temperatures, $T_\delta$ and $T_{60}$, give already a reasonable estimate
of $T_c^0$ for non-zero $H$ and $L^{-1}$.
Forms of universal functions $z_X (z_L)$ along with the optimal parameterized form can be found in
Ref.\cite{Kaczmarek:2020sif}, for the 3-d, $O(4)$ universality class.
Here we present the calculation of $T_c^0$ through $T_\delta$ and $T_{60}$.
Details of the calculational set-up can be found in Ref.\cite{Ding:2019prx}.
Ignoring corrections-to-scaling and keeping in $f_{sub}$ only the leading 
$T$-independent, infinite volume regular contribution proportional to $H$,
we then find for the pseudo-critical temperatures \cite{Ding:2019prx},
\begin{eqnarray}
T_X(H,L)= 
T_c^{0} \left( 1+ \left( \frac{z_X(z_L)}{z_0} \right) 
H^{1/\beta\delta} \right)
+c_X H^{1-1/\delta+1/\beta\delta},\;\; X=\delta,\ 60
 \; .
\label{TX}
\end{eqnarray}

\begin{figure*}[!t]
\centering
\includegraphics[width=0.40\textwidth]{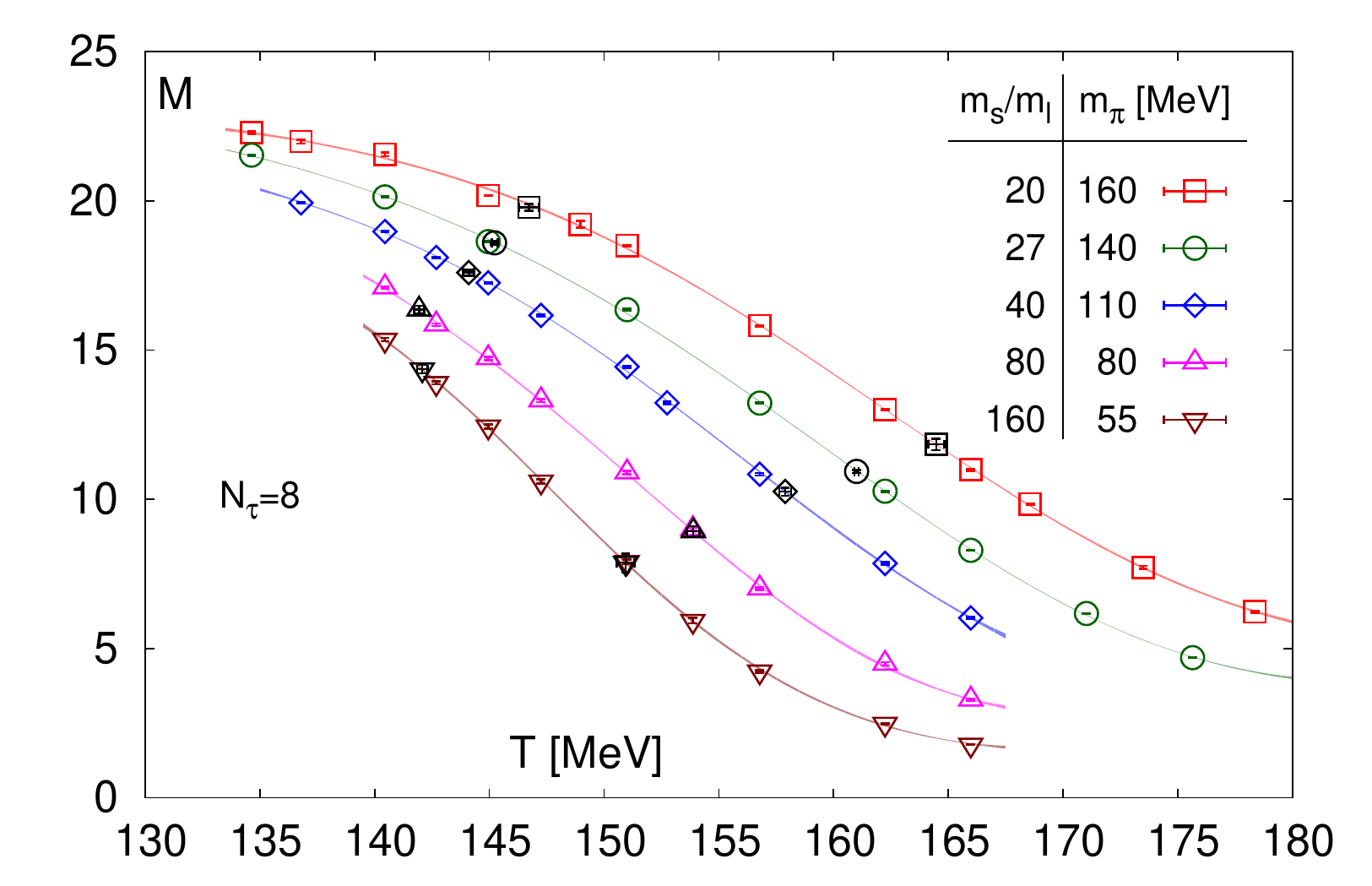} ~~
\includegraphics[width=0.40\textwidth]{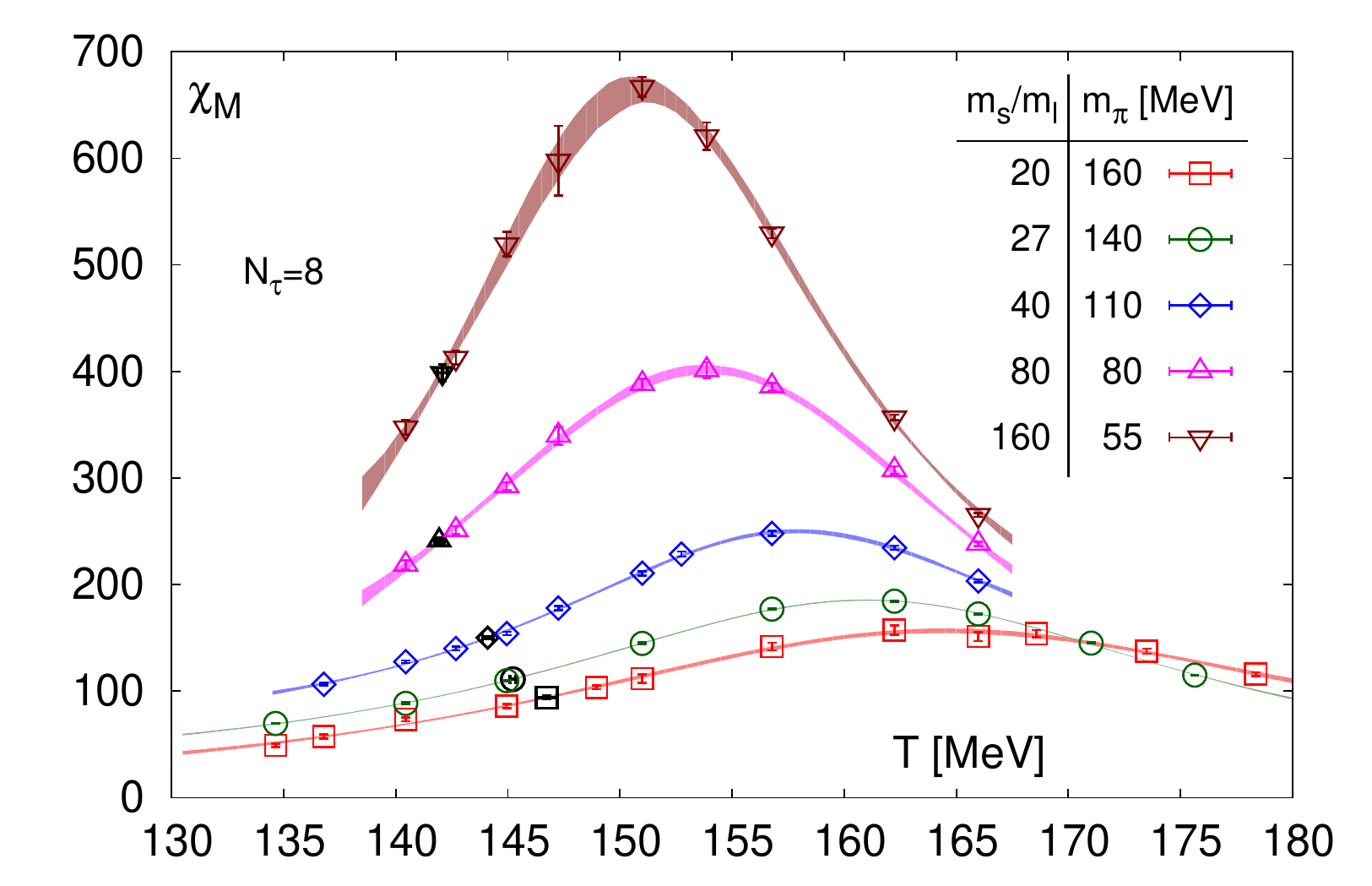}
\caption{
Right: Quark mass dependence of chiral condensate (left) and
the chiral susceptibility (right) on lattices with temporal extent
$N_\tau=8$ for several values of the light quark masses.
The spatial lattices extent $N_\sigma$ is
increased as the light quark mass decreases: $N_\sigma= 32$~
$(H^{-1}=20,\ 27)$, 40~$(H^{-1}=40)$, 56~$(H^{-1}=80,\ 160)$.
Black symbols in the susceptibility plot mark the points corresponding
to 60\% of the peak height and the corresponding values of condensate
has also been marked with black symbols for low temperatures and
the condensates at the peak position of the susceptibility have
also been depicted by bunch of black points at higher temperatures.
Right panel plot is taken from \cite{Ding:2019prx}.
}
\label{fig:massvar}
\end{figure*}

Here we also present the results for the ratio of $M$ and $\chi_M$ and we argue that this ratio
can be used to study the difference between the $O(N)$ and $Z_2$ universality classes, through a non-parametric comparison.
In the infinite volume limit, following Eq.~\ref{scalingexpect} one can write:
\begin{equation}
  \left.\frac{M}{\chi_M}\right|_{T_X,H}=\left.\frac{f_G}{f_{\chi}}\right|_{z_X}H + c_rh_0^{1/\delta}H^{2-1/\delta}
  \left.\left[\frac{f_G}{f_{\chi}}\left(\frac{1}{f_G}-\frac{1}{f_{\chi}}\right)\right]\right|_{z_X}
 \label{eq.univclass}
\end{equation}
where the first term is the universal part and the second term is a regular contribution which
has been calculated by taking $f_{sub}=c_rH$ in Eq.~\ref{scalingexpect}.
We will calculate the ratio in the LHS of Eq.~\ref{eq.univclass} in the thermodynamic limit at different temperatures
like $T_p$ and $T_{60}$ and compare with that from scaling expectation of RHS, where $f_G(z)$ and $f_{\chi}(z)$
will be numbers fixed by the universality class. As can be seen from Eq.~\ref{eq.univclass} a comparison without
the regular term is parameter free. If a second order $Z_2$ endpoint exists at some finite quark mass, $H_c$,
then we have to replace $H$ by $H-H_c$ in the RHS of Eq.~\ref{eq.univclass}.

\section{Results}

We start with the results for $M$ for different values of $H$ which is shown
in Fig.~\ref{fig:massvar} (left) for lattices of size $N_\sigma^3\times N_\tau$
with $N_\tau=8$. One can
see clearly that $M$ decreases with decreasing $H$ and the crossover
becomes sharper towards smaller $H$.
In Fig.~\ref{fig:massvar} (right) we show the chiral susceptibility for lattices
as for Fig.~\ref{fig:massvar} (left). The apparent increase
of peak height is visible with decreasing $H$ and this is consistent with the expected behavior,
$\chi_M^{max} \sim H^{1/\delta-1}+ \text{const}.$, with  $\delta \simeq 4.8$,
although within rather large uncertainty which restricts a precise
determination of $\delta$.

\begin{figure*}[!t]
\centering
\includegraphics[width=0.40\textwidth]{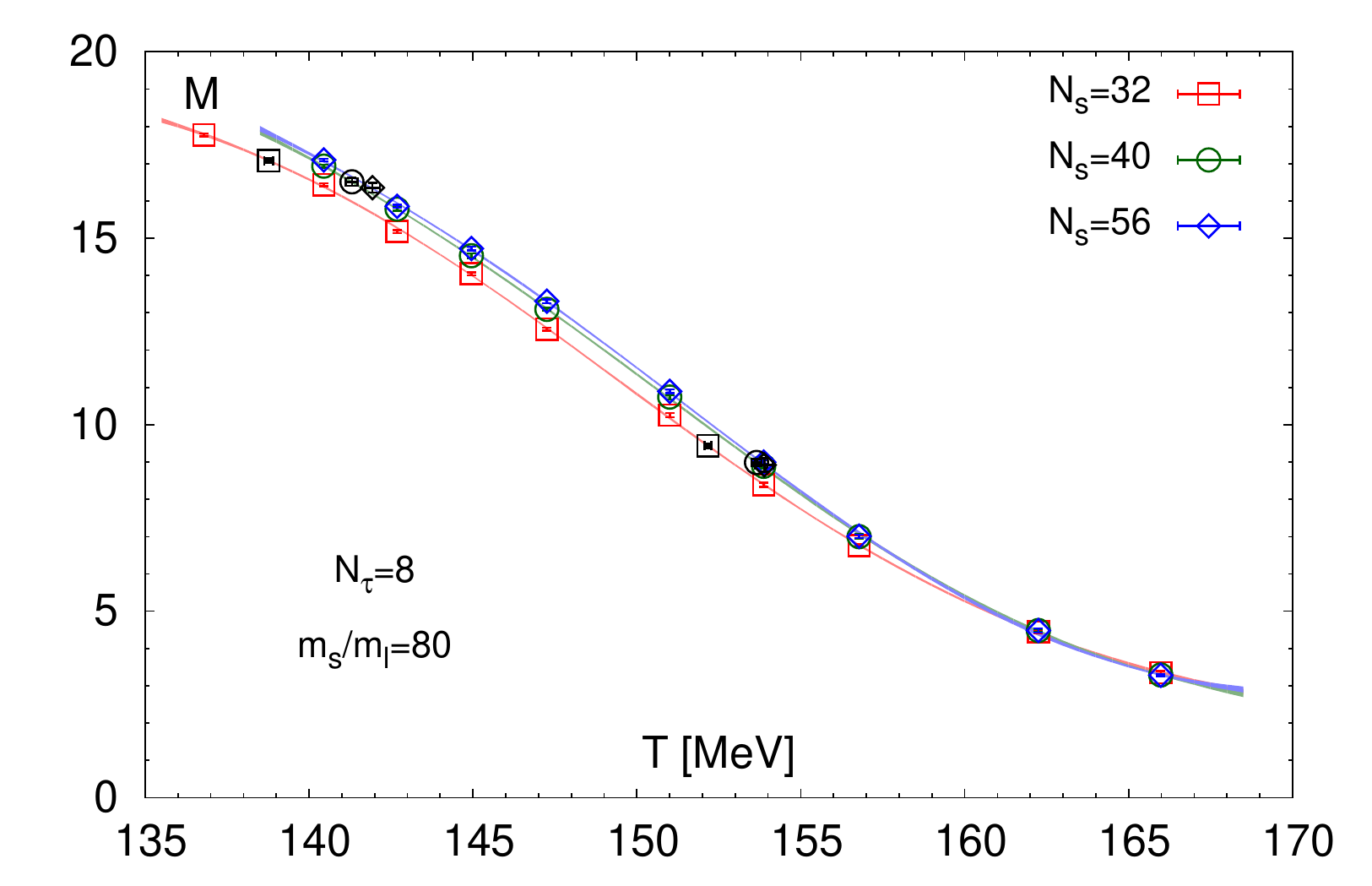} ~~
\includegraphics[width=0.40\textwidth]{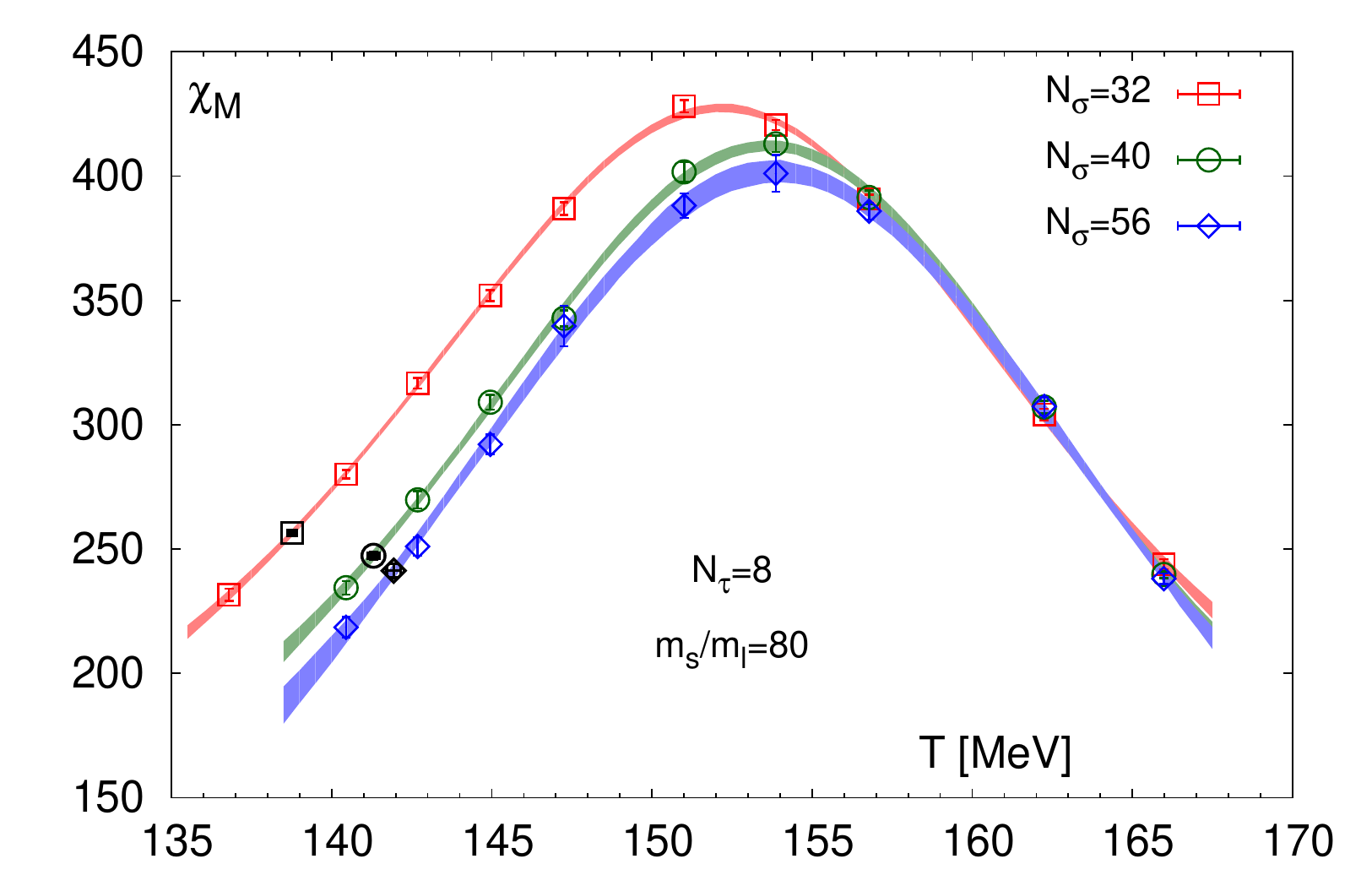}
\caption{
Left: Volume dependence of
the chiral condensate on lattices with temporal extent
$N_\tau=8$ for three different spatial
lattice sizes at $H=1/80$. Bunch of three black symbols at higher and lower
temperatures denote $M$ at $T_p$ and for $T_{60}$, respectively.
Right: Volume dependence of
the chiral susceptibility on lattices with temporal extent
$N_\tau=8$ for three different spatial
lattice sizes at $H=1/80$. Black symbols mark the points corresponding
to 60\% of the peak height. 
Right panel plot is taken from \cite{Ding:2019prx}.
}
\label{fig:volvar}
\end{figure*}

In Fig.~\ref{fig:volvar} (left) we show the volume dependence of the order parameter
for $H=1/80$ on lattices with $N_\tau=8$ and for different aspect
ratios, $N_\sigma/N_\tau=4,\ 5$ and $7$. One can see that $M$ increases
and saturates when approaching the thermodynamic limit. This is found as a basic feature
in $O(4)$ finite size scaling studies \cite{Engels:2014bra} when there is a second order
phase transition for vanishing external field.
In Fig.~\ref{fig:volvar} (right) we show the volume dependence for
the same lattices as for Fig.~\ref{fig:volvar} (left).  
Similar results have also been obtained for $N_\tau=6$ and $12$.
It is important to note that $\chi_M^{max}$ decreases slightly with increasing volume,
contrary to what one would expect to find at or close to a $1^{st}$
order phase transition. In fact this trend also seems to be consistent with the
behavior seen for $O(4)$ universality class finite-size scaling functions \cite{Engels:2014bra}.
Our current results, thus, suggest a continuous phase transition at $H_c=0$.

\begin{figure*}[!t]
\centering
\includegraphics[width=0.30\textwidth]{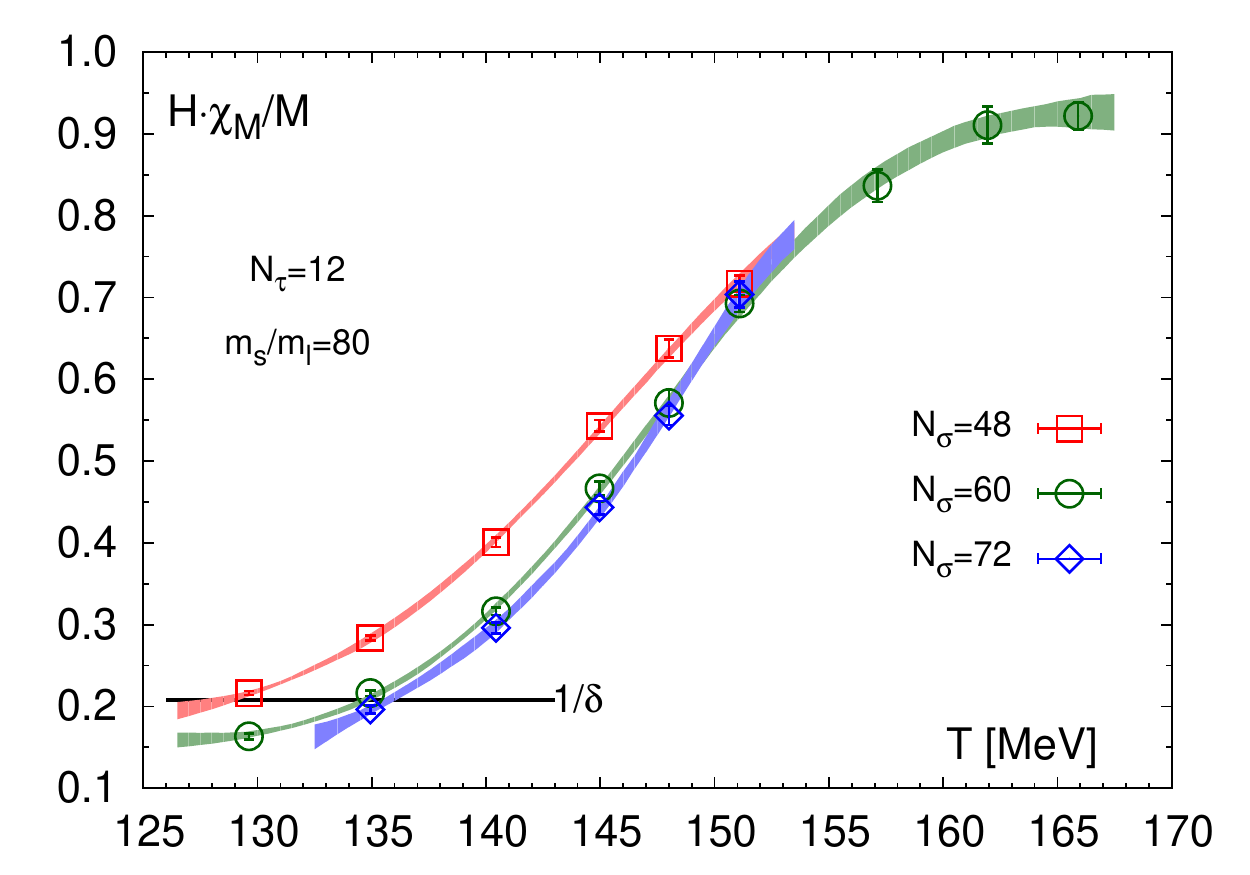} ~~
\includegraphics[width=0.30\textwidth]{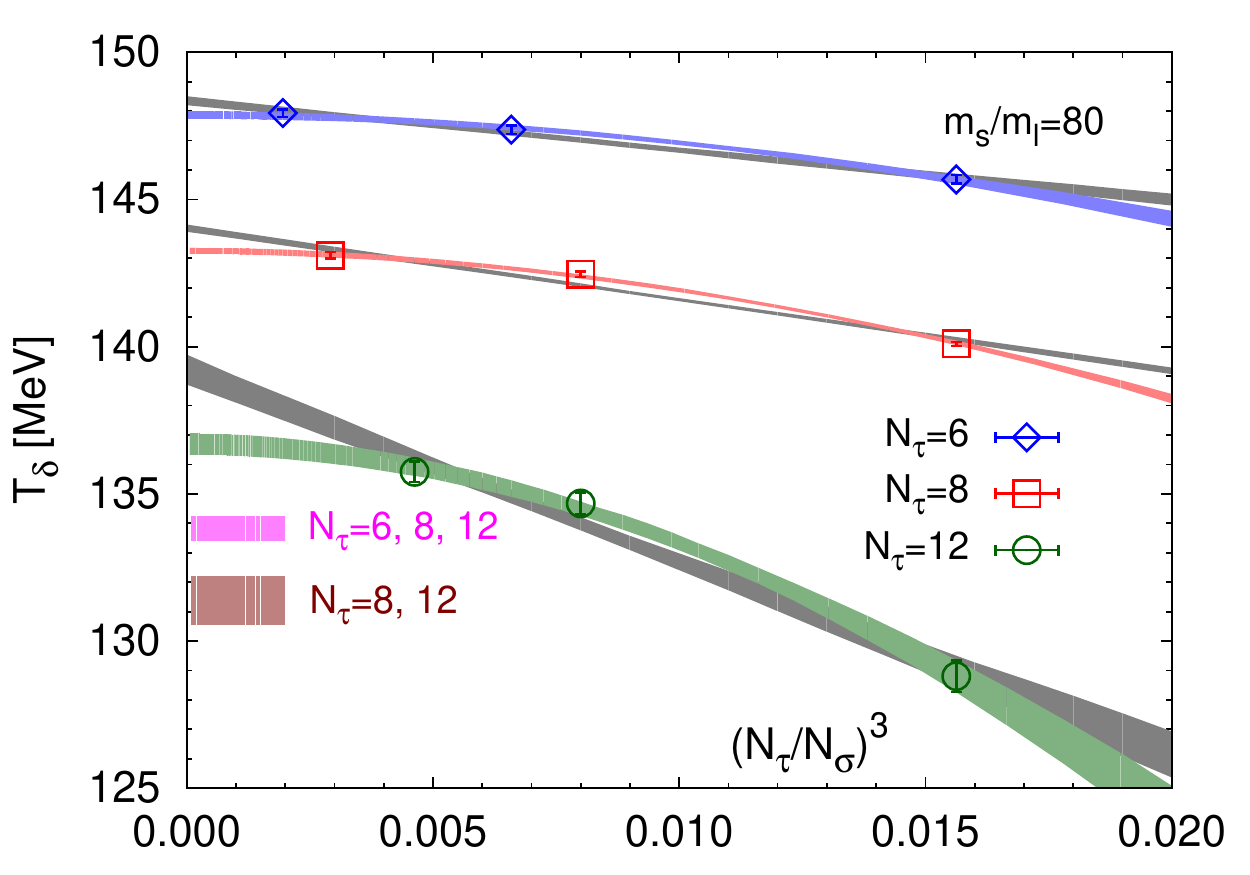} ~~
\includegraphics[width=0.30\textwidth]{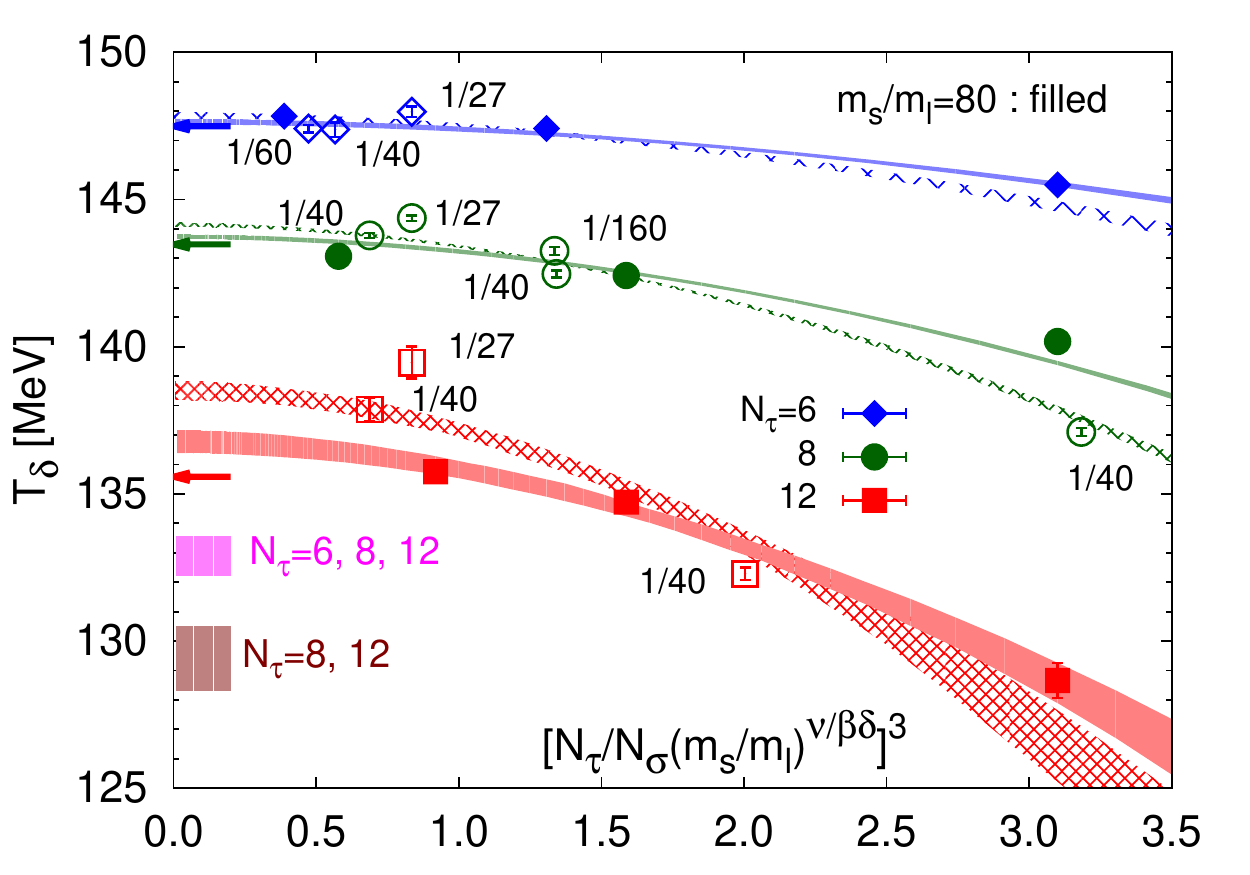}
\caption{
Left: The ratio $H\chi_M/M$ versus temperature for $N_\tau=12$,
$m_l/m_s=1/80$ and different spatial volumes.
Middle: Infinite volume extrapolations based on an $O(4)$ finite-size scaling
function (colored bands) and fits linear in $1/V$ (grey bands). Horizontal
bars show the continuum extrapolated results for $H=1/80$.
Right: Finite size scaling fits for $T_\delta$ based on all
data for $H\le 1/27$ and all available volumes. Arrows show chiral limit
results at fixed $N_\tau$ and horizontal
bars show the continuum extrapolated results for $H=0$.
Figures are taken from \cite{Ding:2019prx}.
}
\label{fig:ratio}
\end{figure*}

Using the results of $M$ and $\chi_M$ we have constructed the ratio $H\chi_M /M$
for lattices with different spatial extents and several values of the light quark masses.
In Fig.~\ref{fig:ratio} (left) this ratio has been shown
for the $N_\tau=12$ lattices with $H=1/80$, as a typical example.
Colored bands are interpolations to the data and the crossings with the
horizontal line at $1/\delta$ define $T_\delta(H,L)$. We extrapolate
$T_\delta(H,L)$ for a fixed $H$ by two methods: (1) using the $O(4)$ scaling
function following Eq.~\ref{TX} where the volume correction is roughly $1/V^2$ \cite{Kaczmarek:2020sif} and
(2) assuming $1/V$ correction which is the case if the volume correction is of regular origin.
The resulting volume extrapolations are shown in Fig.~\ref{fig:ratio} (middle) where it can be seen that
the data seems to reach the thermodynamic limit faster than $1/V$.
The difference between these two extrapolations to the thermodynamic limit serves
as one component of the systematic uncertainty. The same procedure has been followed for
all three different $N_{\tau}$ values and then the continuum extrapolation
of these infinite volume results is performed with and without $N_{\tau}=6$ 
which gives a second source
to the systematic uncertainty. These continuum extrapolated results are shown in
Fig.~\ref{fig:ratio} (middle) by horizontal bands with different colors.
The same set of analyses have also been performed for $H=1/40$.
Finally, we extrapolate the $T_\delta(H,\infty)$ for $H=1/40$
and $1/80$ to the chiral limit using Eq.~\ref{TX}, putting $z_\delta(0)=0$.
Results obtained from these extrapolation chains, with the thermodynamic limit
results
obtained either through $O(4)$ or $1/V$ ansatz, and continuum limit extrapolations with
and without $N_\tau=6$, lead to chiral phase transition temperatures $T_c^0$ 
in the range ($128$-$135$)~MeV.

\begin{wrapfigure}{r}{5.3cm}

\centering
\includegraphics[width=0.45\textwidth]{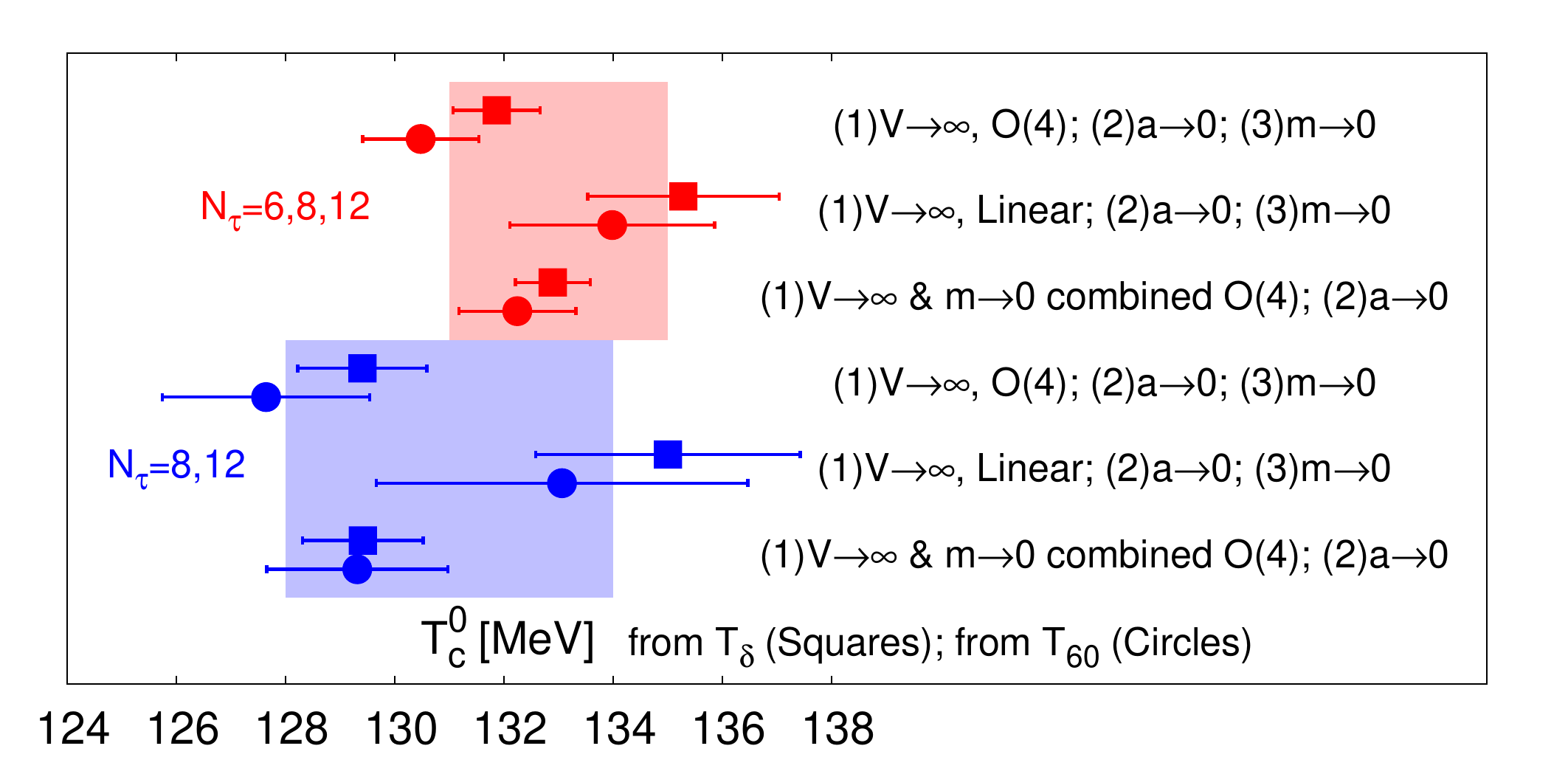}
\caption{Summary of fit results. The order of different limits taken,
described in the main text, is written beside each pair of closest points.
Figure is taken from \cite{Ding:2019prx}.
}
\label{fig:final}

\end{wrapfigure}

Since from Fig.~\ref{fig:ratio} (middle) we can see that the $O(4)$ scaling ansatz is
already working quite well for finite lattice spacing, we attempt a joint extrapolation
to the chiral and thermodynamic limit using results for all masses and on all
available volumes, through $O(4)$ finite-size scaling function.
The resulting extrapolations for three different $N_{\tau}$ are shown in
Fig.~\ref{fig:ratio} (right) where we show the extrapolation only for $H=1/40$ and $H=1/80$
for better readability of the plot. Noticeably for $H=1/80$, these bands compare well 
with the fits shown in Fig.~\ref{fig:ratio} (middle). Colored arrows show the chiral limit
results for each $N_{\tau}$, in the thermodynamic limit. As a final step, the
continuum extrapolation
has been performed again with and without the $N_{\tau}=6$ result. The results 
are shown by the horizontal bars in different colors in Fig.~\ref{fig:ratio} (right).
Results for $T_c^0$, obtained by this method, are also shown in 
Fig.~\ref{fig:final} and they are found to be in
complete agreement with the corresponding numbers when the continuum limit has been taken
before the chiral limit.

For $T_{60}$ same set of analyses has been done as for $T_{\delta}$ and, 
as can be seen
from Fig.~\ref{fig:final}, the resulting $T_c^0$ numbers from analyses of $T_{60}$ agree within 1\%
with the same obtained from $T_{\delta}$ analyses. Leaving out $N_{\tau}=6$ numbers
systematically gives a 2-3 MeV decrease of $T_c^0$, reflected in the displacement of the two
colored bands in Fig.~\ref{fig:final}. Out of all the above-mentioned analyses we finally
quote the chiral phase transition temperature,
\begin{equation}
T_c^0 = 132^{+3}_{-6}~{\rm MeV} \; .
\label{Tcfinal}
\end{equation}

\begin{figure}[!h]
\begin{center}
\includegraphics[width=0.45\textwidth]{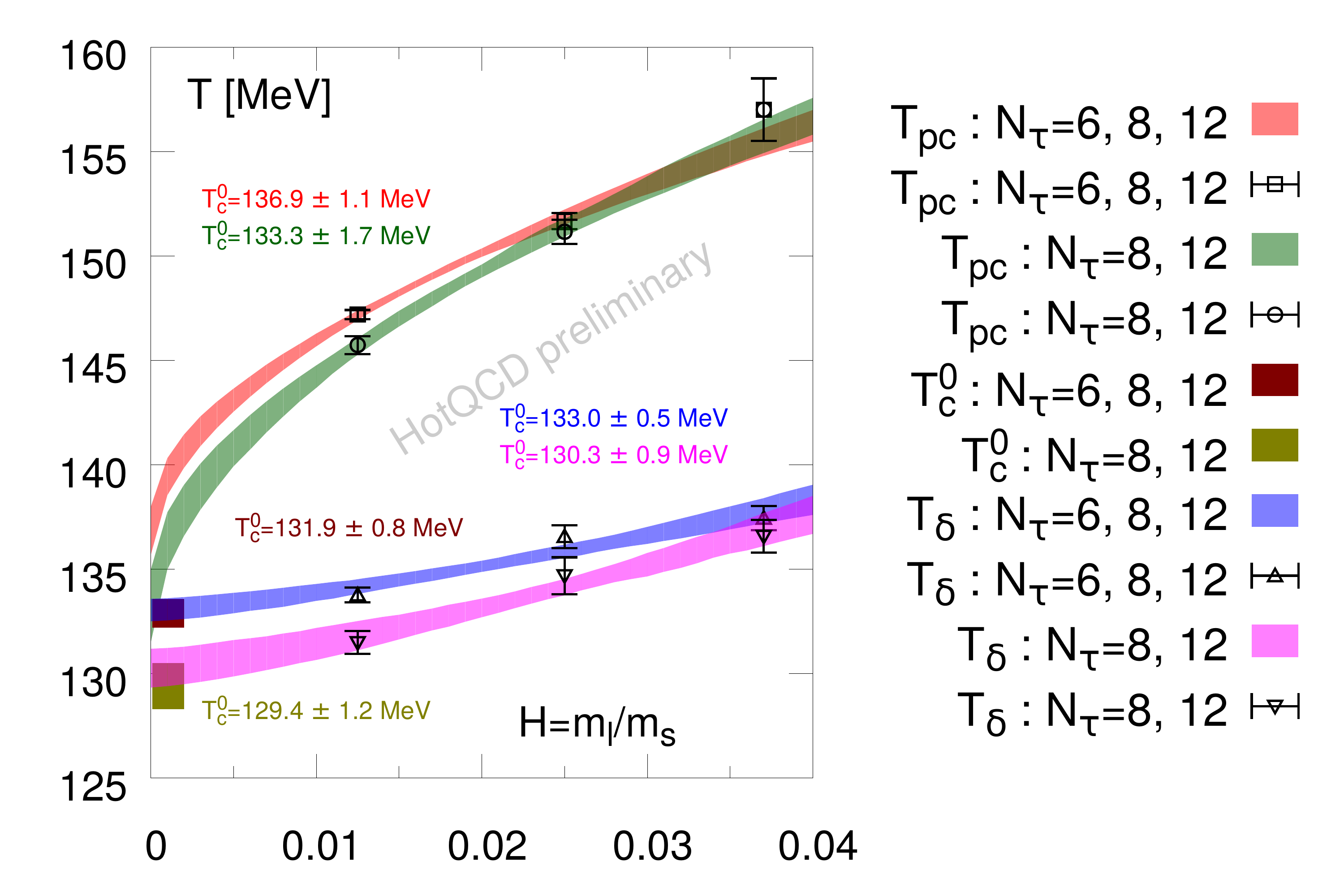} ~~~~
\includegraphics[width=0.45\textwidth]{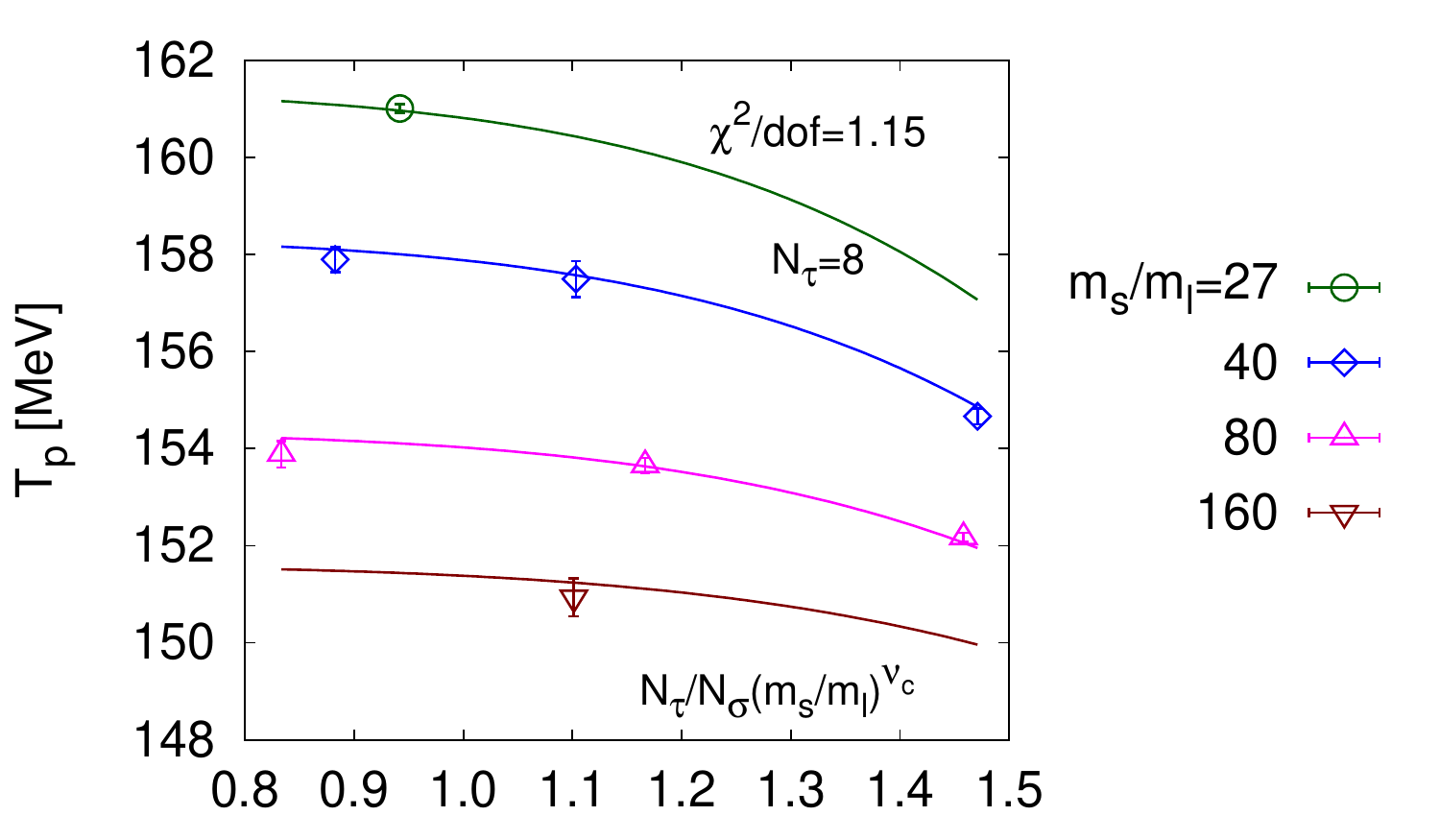}
\end{center}
\caption{Left: Comparison of chiral extrapolation of $T_\delta$ and $T_p$. Results for $H=1/27$
is from lattices with aspect ratio 4 only. Horizontal bands represent the $T_c^0$ obtained
from the $T_\delta$ analyses \cite{Ding:2019prx} without including $H=1/27$.
Right: Joint chiral and infinite volume extrapolation of $T_p$ which gives
$T_c^0=145.6(3)$ MeV for $N_\tau=8$.
}
\label{fig:massextrapolationcompare}
\end{figure}

For completeness we also carried out the analyses for the peak position of the chiral susceptibility,
{\it i.e.}\ $T_p$. Here also we first take the thermodynamic limit and then 
the continuum limit.
Results for continuum extrapolated $T_p(H,\infty)$ are shown in Fig.~\ref{fig:massextrapolationcompare} (left).
For the extrapolation of $T_p$ we could not include the sub-leading contributions, which is
of course, important. Including such a term with the three $H$ values at our disposal, makes the chiral
extrapolation way less controlled. Apparently the $T_c^0$ obtained from the extrapolations of $T_p$,
even without a regular contribution, are in agreement with the same from $T_\delta$ within 95\% confidence.
The inclusion of a regular term will presumably make the agreement even better. 
Inclusion of $N_\tau=6$ results, as usual gives systematically higher $T_p$,
similar to the case of $T_\delta$, as mentioned earlier.
The numbers in Fig.~\ref{fig:massextrapolationcompare} (left) are preliminary.
Here we also show the $T_p$ and $T_\delta$ for $H=1/27$,
for lattices with aspect ratio 4 only, for which finite volume effects have 
been estimated to be 
similar to the magnitude of the present uncertainty. This can be seen from
Fig.~\ref{fig:massextrapolationcompare} (right) where we have shown the joint chiral and
thermodynamic limit extrapolation for $N_\tau=8$ using the $O(4)$ finite size scaling functions.
A similar analysis has also been carried out for $N_\tau=6$ and $12$. A continuum extrapolation
then gives $T_c^0$ consistent with Fig.~\ref{fig:massextrapolationcompare} (left) and
other estimates within 95\% confidence.

\begin{figure}[!h]
\centering
\includegraphics[width=0.45\textwidth]{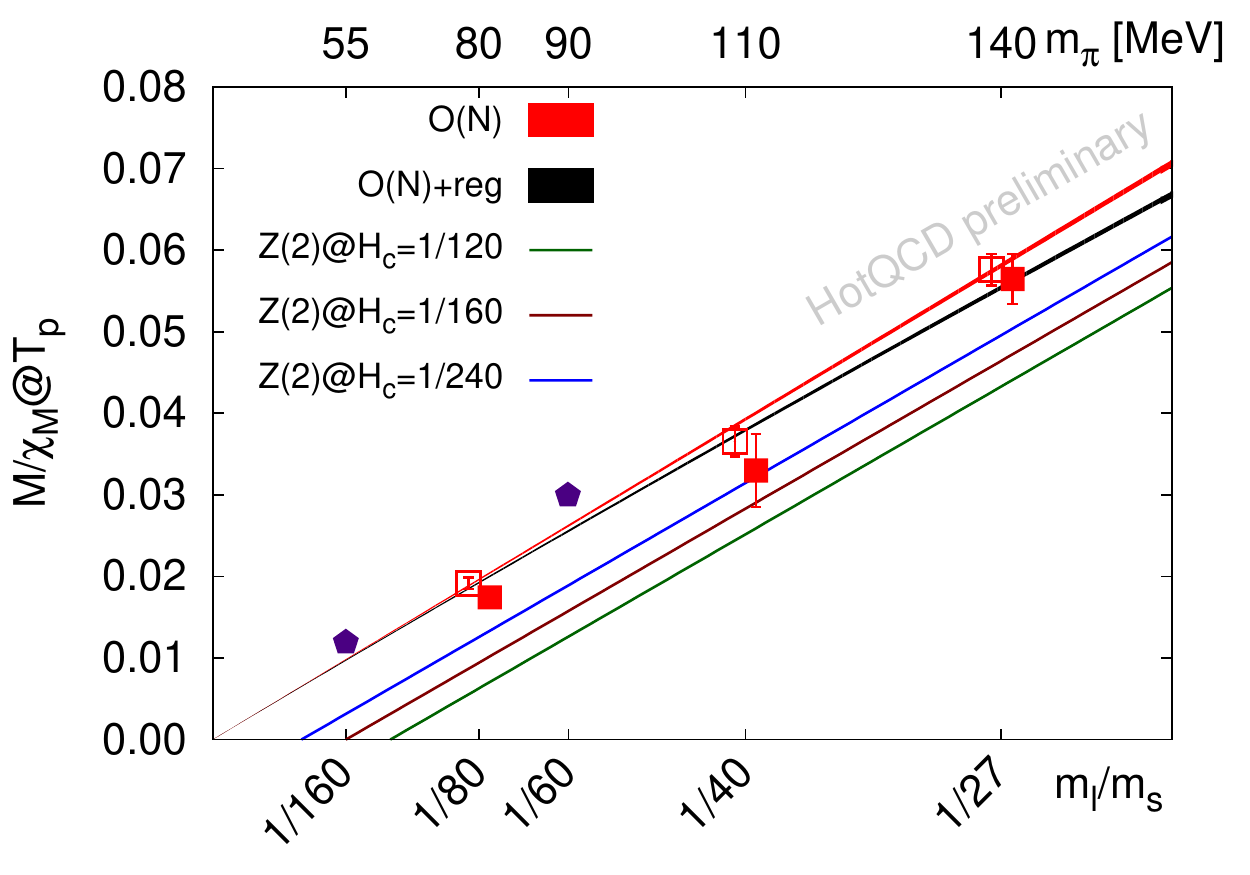} ~~
\includegraphics[width=0.45\textwidth]{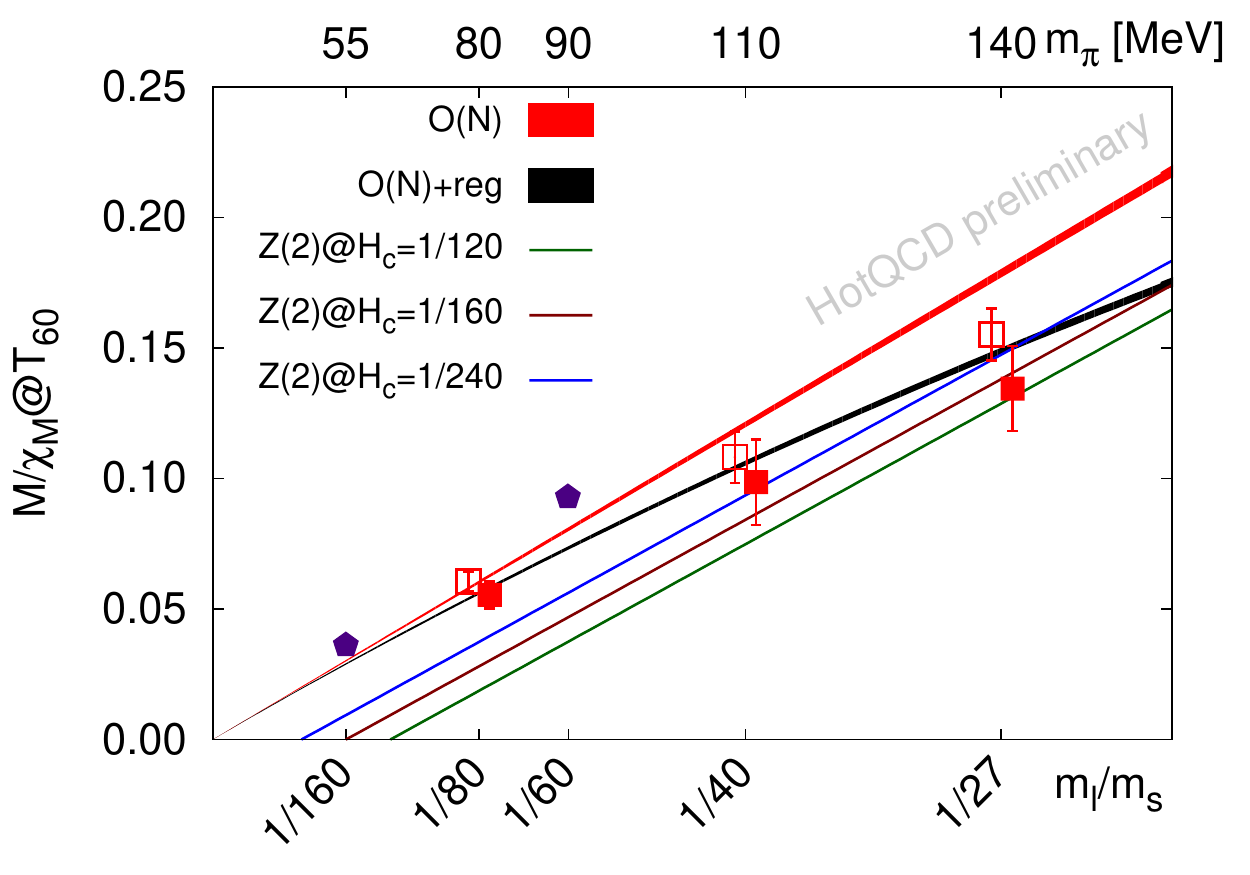}
\caption{Mass scaling of the ratio of $M$ and $\chi_M$ at the peak (left)
and at the 60\% of the peak (right). The bands (hardly distinguishable from
thick lines) are not fits and the width
of the bands are due to the difference between $O(4)$ and $O(2)$ and for that
reason the band collectively represented as $O(N)$. The squares represent
the continuum extrapolated (open: with $N_\tau=6$ and filled: without $N_\tau=6$)
results and the pentagons represent results for $N_\tau=6$ and 8 with
$H=1/60$ and 1/160, respectively.
}
\label{fig:univclass}
\end{figure}

Next we show  in Fig.~\ref{fig:univclass} left and right results for the 
ratio of $M$ and $\chi_M$ evaluated at different temperatures,
{\it i.e.}\ $T_{p}$ and $T_{60}$, respectively.
As discussed earlier this ratio gives a handle to compare the QCD results with 
results for different
universality classes in a non-parametric way. We first calculate the ratio $M/\chi_M$ at the specified
temperatures on different volumes for a fixed mass and fixed $N_\tau$.
We have already seen in Fig.~\ref{fig:volvar} (left) and Fig.~\ref{fig:volvar} (right)
that the volume dependences of $M$ and $\chi_M$ are very small both at $T_p$
and $T_{60}$. As a result the volume dependence of the ratio $M/\chi_M$ is also found to
be small and in most of the cases a linear volume extrapolation gives the thermodynamic
limit result which is in good agreement with that from the largest volume 
available.
As usual for $H=1/27$ we show the result for aspect ratio 4 and for this ratio
we did not apply estimated correction because of the above-mentioned reason. Next we continuum
extrapolate the ratio for a fixed $H$ and these continuum extrapolated results are depicted
in Fig.~\ref{fig:univclass}. For $H=1/60$ and 1/160 we have results only for $N_\tau=6$ and 8, respectively.
Like in other cases we checked the systematic uncertainty in the continuum extrapolation
by including and excluding $N_\tau=6$ results. In Fig.~\ref{fig:univclass} we show the scaling
expectations for $O(4)$ (relevant for continuum extrapolated cases) and $O(2)$ (relevant for results obtained with
a finite lattice spacing) universality classes following Eq.~\ref{eq.univclass}.
Expectations based on both universality classes differ little and 
are plotted  together as a band denoted as $O(N)$. For the regular part of Eq.~\ref{eq.univclass}
we did not fit the coefficient $c_r$ from the $M/\chi_M$ ratio. For our preliminary comparison we rather took the values
of $c_r$ and $h_0$ from the fit of $M$ and $\chi_M$. This
seems to describe the data quite satisfactorily up to physical masses.
In Fig.~\ref{fig:univclass} one can see that the effect of a regular term is more
important at $T_{60}$ compared to $T_p$ which seems to be counter-intuitive.
Since the contribution of a ($T$-independent) regular term compared to the
singular contribution rises for $M$ and decreases for $\chi_M$ when one goes
from $T_{60}$ to $T_p$, a depreciation of the regular contribution of the ratio
$M/\chi_M$ happens, which can be realized by looking at Eq.~\ref{eq.univclass}.
We also show the scaling expectations for
$Z_2$ universality class for different values of $H_c$. Of course, when there is a $Z_2$
endpoint at some non-vanishing $H_c$, then $M$, defined in Eq.~\ref{pbp} is not an exact
order parameter anymore \cite{Karsch:2000xv,Karsch:2001nf}. Although one has to keep in
mind that the mixing between magnetization like and energy like operators 
becomes smaller
when $H_c$ decreases. Fig.~\ref{fig:univclass} apparently shows that with the current
calculations, the existence of a $Z_2$ endpoint is unlikely down to $H_c=1/240$ corresponding
to $m_\pi\sim 45$ MeV.

\section{Conclusions}
Based on two novel estimators, we have calculated the chiral
phase transition temperature in QCD with two massless light quarks and a
physical strange quark. Eq.~\ref{Tcfinal} lists our thermodynamic-, continuum- and
chiral-extrapolated result for the chiral phase transition temperature,  which is
about $25$~MeV smaller than the pseudo-critical (crossover)  temperature,
$T_{pc}$ for physical values of the light and strange quark masses.
Preliminary calculation of $T_c^0$ from the peak positions gives results
which are in agreement within 95\% confidence with the results from new
estimators. We also showed the results for the ratio of the chiral condensate and
chiral susceptibility which we used to differentiate between $O(N)$ and $Z_2$
universality classes in a non-parametric manner and we found that the
existence of second order endpoint belonging to $Z_2$ universality
class seems to be unlikely down to $m_\pi\sim 45$ MeV.

\section{Acknowledgement}
This work was supported by
the Deutsche Forschungsgemeinschaft (DFG) through Grant No. 315477589-TRR 211 and by
Grant No. 05P18PBCA1 of the German Bundesministerium f\"ur Bildung und Forschung.

\end{document}